\documentclass{article}
\usepackage{amsfonts}
\usepackage{amsmath}
\usepackage{hyperref}

\input{tcilatex}
\begin{document}

\title{Charge Densities for Conducting Ellipsoids}
\author{T L Curtright$^{1}$, Z Cao, S Huang, J S Sarmiento,\smallskip \and S
Subedi, D A Tarrence, and T R Thapaliya\bigskip \\
%EndAName
Department of Physics, University of Miami, Coral Gables, FL 33124-8046, USA}
\maketitle

\begin{abstract}
The volume charge density for a conducting ellipsoid is expressed in simple
geometrical terms, and then used to obtain the known surface charge density
as well as the uniform charge per length along any principal axis. \
Corresponding results are presented for conducting hyperellipsoids in any
number of spatial dimensions.
\end{abstract}

\section{Introduction}

The electrostatics of charged conducting ellipsoids embedded in three
dimensions were first understood in the early part of the nineteenth century 
\cite{Green,Murphy}. \ The surface charge densities as well as the
potentials and electric fields surrounding such objects have elegant
geometrical properties, as discussed extensively in the literature \cite%
{Kelvin,Routh,Kellogg,Sommerfeld,Smythe,Durand}. \ In particular, the
equipotentials are confocal ellipsoids surrounding the charged surface, with
the electric field everywhere normal to those ellipsoids. \ More
specifically, the electric field at any observation point outside a charged
conducting ellipsoid of revolution, either prolate or oblate, is always
directed along the bisector of a pair of straight lines drawn from either of
two focal points of the ellipsoid to the observation point. \ Moreover, for
any conducting ellipsoid in three spatial dimensions, the charge per length
is always constant when projected along any of the three principal axes, a
feature that is perhaps the one elementary property that is easiest to keep
in mind. \ However, in any other number of spatial dimensions, this last
statement must be modified, as discussed recently in \cite{TLC}.%
\footnotetext[1]{%
curtright@miami.edu}

In any case, it may \emph{not} be so well-known that the surface and linear
charge densities for conducting ellipsoids follow easily from elementary 
\emph{volume} charge densities that are simply expressed in geometrical
terms using Dirac deltas. \ The purpose of this article is to present the
volume charge densities and obtain from these the known surface and linear
densities, in any number of spatial dimensions. \ For completeness, we also
discuss the potentials and electric fields surrounding conducting
ellipsoids, thereby confirming the surface charge density through the use of
Gauss' law. \ Finally, an Appendix discusses the geometry of hyperellipsoids
from both intrinsic and extrinsic points of view.

\section{Charge densities for ellipsoids in three dimensions}

Let us begin with the charge distribution on an ideal, static, conducting
ellipsoidal surface embedded in three spatial dimensions. \ The charged
surface is specified by%
\begin{equation}
\frac{x^{2}}{a^{2}}+\frac{y^{2}}{b^{2}}+\frac{z^{2}}{c^{2}}=1\ ,
\label{Ellipsoid3D}
\end{equation}%
where we have expressed the constraint that defines the ellipsoid in the
most convenient Cartesian frame. \ If this surface is an ideal conductor,
i.e. an equipotential surface, carrying a total static charge $Q$, then that
charge is distributed according to a \emph{volume} charge density given
simply by\footnote{%
Note that integration over all space immediately gives $\int \rho \left( 
\overrightarrow{r}\right) dxdydz=Q$ just by rescaling $x,y,z\rightarrow
ax,by,cz$.}%
\begin{equation}
\rho \left( \overrightarrow{r}\right) =\frac{Q}{4\pi abc}~\delta \left( 
\sqrt{\frac{x^{2}}{a^{2}}+\frac{y^{2}}{b^{2}}+\frac{z^{2}}{c^{2}}}-1\right)
\ .  \label{Rho3D}
\end{equation}%
The Dirac delta $\delta \left( \sqrt{\frac{x^{2}}{a^{2}}+\frac{y^{2}}{b^{2}}+%
\frac{z^{2}}{c^{2}}}-1\right) $ restricts all the charge to lie on the
surface, albeit not uniformly if any two of $a,b,c$ are unequal, and is an
obvious property of $\rho $ for this system. \ On the other hand, it is
remarkable and perhaps surprising that the coefficient of this $\delta $ is
a \emph{constant}. \ Of course, when $a=b=c$ the expression (\ref{Rho3D})
reduces to the well-known $\rho $ for a uniformly charged, hollow spherical
shell of radius $a$. \ 

Nevertheless, although this form for $\rho \left( \overrightarrow{r}\right) $%
\ is remarkably simple even when $a$, $b$, and $c$ are all different, we are 
\emph{not} aware of any previous literature that gives the explicit result (%
\ref{Rho3D}). \ For a complete justification of (\ref{Rho3D}), we next
compute from $\rho \left( \overrightarrow{r}\right) $ the corresponding
surface charge density $\sigma \left( \overrightarrow{r}\right) $ on the
general triaxial ellipsoid defined by (\ref{Ellipsoid3D}).

Our point is just that, for a volume density on a surface defined by $%
F\left( \overrightarrow{r}\right) =0$, of the form 
\begin{equation}
\rho \left( \overrightarrow{r}\right) =f\left( \overrightarrow{r}\right)
~\delta \left( F\left( \overrightarrow{r}\right) \right) \ ,
\label{GenericForm}
\end{equation}%
\emph{the restriction of the function} $f\left( \overrightarrow{r}\right) $ 
\emph{to the surface} \emph{is uniquely determined by the surface charge
density} $\sigma \left( \overrightarrow{r}\right) $, and vice versa. \ It is
only necessary to integrate $\rho \left( \overrightarrow{r}\right) $ along a
line normal to the charged surface to obtain $\sigma \left( \overrightarrow{r%
}\right) $, and thereby determine $f\left( \overrightarrow{r}\right) $. \ It
follows from a straightforward calculation that the relation between $f$ and 
$\sigma $ is given by%
\begin{equation}
f\left( \overrightarrow{r}\right) =\left\vert \overrightarrow{\nabla }%
F\left( \overrightarrow{r}\right) \right\vert \sigma \left( \overrightarrow{r%
}\right)
\end{equation}%
for points $\overrightarrow{r}$\ on the surface. \ To obtain the known
surface charge density for a triaxial ellipsoid from (\ref{Rho3D}), for
arbitrary $a$, $b$, and $c$, the calculation goes as follows. \ 

The normal unit vector at any point \emph{on the surface} is given by%
\begin{equation}
\widehat{n}=\frac{\overrightarrow{\nabla }\left(
x^{2}/a^{2}+y^{2}/b^{2}+z^{2}/c^{2}\right) }{\left\vert \overrightarrow{%
\nabla }\left( x^{2}/a^{2}+y^{2}/b^{2}+z^{2}/c^{2}\right) \right\vert }=%
\frac{x~\widehat{x}/a^{2}+y~\widehat{y}/b^{2}+z~\widehat{z}/c^{2}}{\sqrt{%
x^{2}/a^{4}+y^{2}/b^{4}+z^{2}/c^{4}}}\ ,  \label{Normal3D}
\end{equation}%
while the three-space volume element in an infinitesimal neighborhood
straddling the surface is%
\begin{equation}
dV=du~dA\ ,\ \ \ du=\widehat{n}\cdot d\overrightarrow{r}\ .  \label{dV3D}
\end{equation}%
Rewriting the Dirac delta in terms of the normal coordinate $u$ then gives%
\begin{eqnarray}
\delta \left( \sqrt{\frac{x^{2}}{a^{2}}+\frac{y^{2}}{b^{2}}+\frac{z^{2}}{%
c^{2}}}-1\right) &=&\frac{1}{\left\vert \frac{d}{du}\sqrt{\frac{x^{2}}{a^{2}}%
+\frac{y^{2}}{b^{2}}+\frac{z^{2}}{c^{2}}}\right\vert }~\delta \left(
u-u_{0}\right) =\frac{1}{\left\vert \widehat{n}\cdot \overrightarrow{\nabla }%
\sqrt{\frac{x^{2}}{a^{2}}+\frac{y^{2}}{b^{2}}+\frac{z^{2}}{c^{2}}}%
\right\vert }~\delta \left( u-u_{0}\right)  \notag \\
&=&\frac{1}{\sqrt{\frac{x^{2}}{a^{4}}+\frac{y^{2}}{b^{4}}+\frac{z^{2}}{c^{4}}%
}}~\delta \left( u-u_{0}\right) \ ,
\end{eqnarray}%
where\footnote[3]{%
For a given point on the ellipsoid, the value for $u_{0}$ is unique,
obviously.} $u=u_{0}$\ when $x^{2}/a^{2}+y^{2}/b^{2}+z^{2}/c^{2}=1$. \ Using
this last expression and integrating over the infinitesimal neighborhood $%
u\in \left( u_{0}-\varepsilon ,u_{0}+\varepsilon \right) $ gives the
expected result (e.g. see \cite{Smythe}) for the surface charge density,%
\begin{equation}
\sigma \left( \overrightarrow{r}\right) =\lim_{\varepsilon \rightarrow
0}\int_{u_{0}-\varepsilon }^{u_{0}+\varepsilon }\rho \left( \overrightarrow{r%
}\right) du=\frac{Q}{4\pi abc}\frac{1}{\sqrt{\frac{x^{2}}{a^{4}}+\frac{y^{2}%
}{b^{4}}+\frac{z^{2}}{c^{4}}}}\ ,  \label{Sigma3D}
\end{equation}%
thereby confirming that (\ref{Rho3D}) is correct. \ In this last
expression,\ it is to be understood that all points $\overrightarrow{r}$\
are on the surface (\ref{Ellipsoid3D}).

The volume charge density (\ref{Rho3D}) is also convenient to show that the
projected charge/length along any principal axis is constant. \ For example,
using the Dirac delta property $\delta \left( f\left( z\right) \right)
=\sum_{\text{roots }z_{0}~\text{of }f}~\delta \left( z-z_{0}\right)
/\left\vert f^{\prime }\left( z_{0}\right) \right\vert $, we have for any $x$
between $\pm a$,%
\begin{gather}
\frac{dQ}{dx}\equiv \int_{-\infty }^{+\infty }dy\int_{-\infty }^{+\infty
}dz~\rho \left( \overrightarrow{r}\right) =\frac{Q}{4\pi abc}~\int_{-\infty
}^{+\infty }dy\int_{-\infty }^{+\infty }dz~\delta \left( \sqrt{\frac{x^{2}}{%
a^{2}}+\frac{y^{2}}{b^{2}}+\frac{z^{2}}{c^{2}}}-1\right)  \notag \\
=\frac{Q}{4\pi abc}~\int_{-\infty }^{+\infty }dy\int_{-\infty }^{+\infty
}dz~\left\vert \frac{c^{2}}{z}\right\vert \left[ \delta \left( z-c\sqrt{%
1-x^{2}/a^{2}-y^{2}/b^{2}}\right) +\delta \left( z+c\sqrt{%
1-x^{2}/a^{2}-y^{2}/b^{2}}\right) \right] _{u=u_{0}}  \notag \\
=\frac{Q}{2\pi ab}~\int_{-b\sqrt{1-x^{2}/a^{2}}}^{+b\sqrt{1-x^{2}/a^{2}}}%
\frac{dy}{\sqrt{1-x^{2}/a^{2}-y^{2}/b^{2}}}=\frac{Q}{2\pi a}~\int_{-1}^{+1}%
\frac{ds}{\sqrt{1-s^{2}}}\ .
\end{gather}%
But then $\int_{-1}^{+1}\frac{ds}{\sqrt{1-s^{2}}}=\pi $, so for $-a\leq
x\leq a$, 
\begin{equation}
\frac{dQ}{dx}=\frac{Q}{2a}\ .  \label{Lambda3D}
\end{equation}%
Similarly, the projected linear charge densities along the other principal
axes are given by $dQ/dy=Q/\left( 2b\right) $ for $-b\leq y\leq b$ and $%
dQ/dz=Q/\left( 2c\right) $ for $-c\leq z\leq c$.

\section{Charge densities for hyperellipsoids}

The results and analysis given above are easily extended to describe
equipotential static conducting ellipsoids embedded in any number of spatial
dimensions, $D$. \ These are sometimes called \textquotedblleft
hyperellipsoids\textquotedblright\ for $D>3$. \ At the risk of being
somewhat repetitive, let us consider this generalization in detail.

The corresponding volume charge distribution, $\rho _{D}\left( 
\overrightarrow{r}\right) $, that is appropriate for equipotential
hyperellipsoids in $D$ dimensions, is given by an immediate generalization
of (\ref{Rho3D}), namely,\footnote{%
Note that integration over all space immediately gives $\int \rho \left( 
\overrightarrow{r}\right) dx_{1}dx_{2}\cdots dx_{D}=Q$ just by rescaling $%
x_{1},x_{2},\cdots ,x_{D}\rightarrow a_{1}x_{1},a_{2}x_{2},\cdots
,a_{D}x_{D} $.} 
\begin{eqnarray}
\rho _{D}\left( \overrightarrow{r}\right) &=&\frac{Q}{\Omega
_{D}~a_{1}a_{2}\cdots a_{D}}~\delta \left( \sqrt{\frac{x_{1}^{2}}{a_{1}^{2}}+%
\frac{x_{2}^{2}}{a_{2}^{2}}+\cdots +\frac{x_{D}^{2}}{a_{D}^{2}}}-1\right)
=\sigma _{D}\left( \overrightarrow{r}\right) ~\delta \left( u-u_{0}\right) \
,  \label{RhoAnyD} \\
\sigma _{D}\left( \overrightarrow{r}\right) &=&\frac{Q}{\Omega
_{D}~a_{1}a_{2}\cdots a_{D}}\left. \frac{1}{\sqrt{%
x_{1}^{2}/a_{1}^{4}+x_{2}^{2}/a_{2}^{4}+\cdots +x_{D}^{2}/a_{D}^{4}}}%
\right\vert _{u=u_{0}}\ ,  \label{SigmaAnyD}
\end{eqnarray}%
where $u=u_{0}$\ when $x_{1}^{2}/a_{1}^{2}+x_{2}^{2}/a_{2}^{2}+\cdots
+x_{D}^{2}/a_{D}^{2}=1$ and where\footnote{%
It is interesting to compare the hypersurface charge density $\sigma _{D}$
with the product of the principal curvatures, $\kappa _{m}$, $m=1,2,\cdots
,D-1$, for the same ellipsoidal hypersurface. \ A straightforward
calculation (see the Appendix) gives%
\begin{equation*}
\left( \Omega _{D}\sigma _{D}\right) ^{D+1}\mathbb{=}\left( \dprod_{j=1}^{D}%
\frac{1}{a_{j}^{D-1}}\right) \left( \dprod_{m=1}^{D-1}\kappa _{m}\right)
\end{equation*}%
} 
\begin{equation}
\Omega _{D}=\frac{2\pi ^{D/2}}{\Gamma \left( D/2\right) }\ .  \label{OmegaD}
\end{equation}%
This $\Omega _{D}\ $is the total \textquotedblleft solid
angle\textquotedblright\ in $D$ spatial dimensions. \ The unit normal vector
and the volume measure, for an infinitesimal neighborhood straddling the $%
D-1 $ dimensional \textquotedblleft hypersurface\textquotedblright\
containing the charge, are also given by the obvious generalizations of (\ref%
{Normal3D}) and (\ref{dV3D}). 
\begin{equation}
\widehat{n}=\frac{x_{1}~\widehat{x}_{1}/a_{1}^{2}+x_{2}~\widehat{x}%
_{2}/a_{2}^{2}+\cdots +x_{D}~\widehat{x}_{D}/a_{D}^{2}}{\sqrt{%
x_{1}^{2}/a_{1}^{4}+x_{2}^{2}/a_{2}^{4}+\cdots +x_{D}^{2}/a_{D}^{4}}}\ ,\ \
\ d^{D}V=du~d^{D-1}V\ ,\ \ \ du=\widehat{n}\cdot d\overrightarrow{r}\ .
\label{NormalD}
\end{equation}

The volume charge density (\ref{RhoAnyD}) can be projected along any
principal axis to obtain a \emph{non}-uniform linear charge density for $%
D\neq 3$. \ As done before for $D=3$, the projection is defined by
integrating over all but one direction. \ For instance,%
\begin{equation}
\frac{dQ}{dx_{1}}\equiv \int_{-\infty }^{+\infty }dx_{2}~\cdots
\int_{-\infty }^{+\infty }dx_{D}~\rho _{D}\left( \overrightarrow{r}\right) =%
\frac{Q}{\Omega _{D}~a_{1}a_{2}\cdots a_{D}}~\int_{-\infty }^{+\infty
}dx_{2}\cdots \int_{-\infty }^{+\infty }dx_{D}~\delta \left( \sqrt{\frac{%
x_{1}^{2}}{a_{1}^{2}}+\frac{x_{2}^{2}}{a_{2}^{2}}+\cdots +\frac{x_{D}^{2}}{%
a_{D}^{2}}}-1\right) \ .
\end{equation}%
Now perform the integrations sequentially, beginning with $\int_{-\infty
}^{+\infty }dx_{D}$ using the same property of the Dirac delta as was used
for $D=3$. 
\begin{equation}
\delta \left( \sqrt{\frac{x_{1}^{2}}{a_{1}^{2}}+\frac{x_{2}^{2}}{a_{2}^{2}}%
+\cdots +\frac{x_{D}^{2}}{a_{D}^{2}}}-1\right) =\left\vert \frac{a_{D}^{2}}{%
x_{D}}\right\vert \left[ 
\begin{array}{c}
\delta \left( x_{D}-a_{D}\sqrt{1-x_{1}^{2}/a_{1}^{2}-\cdots
-x_{D-1}^{2}/a_{D-1}^{2}}\right) \\ 
\\ 
+\delta \left( x_{D}+a_{D}\sqrt{1-x_{1}^{2}/a_{1}^{2}-\cdots
-x_{D-1}^{2}/a_{D-1}^{2}}\right)%
\end{array}%
\right] _{u=u_{0}}\ .
\end{equation}%
This leads to the next integration,%
\begin{equation}
\frac{2}{a_{D-1}}\int_{-a_{D-1}\sqrt{1-x_{1}^{2}/a_{1}^{2}-\cdots
-x_{D-2}^{2}/a_{D-2}^{2}}}^{+a_{D-1}\sqrt{1-x_{1}^{2}/a_{1}^{2}-\cdots
-x_{D-2}^{2}/a_{D-2}^{2}}}dx_{D-1}~\frac{1}{\sqrt{1-x_{1}^{2}/a_{1}^{2}-%
\cdots -x_{D-1}^{2}/a_{D-1}^{2}}}=\frac{2\pi }{a_{D-1}}\ ,
\end{equation}%
and therefore%
\begin{equation}
\frac{dQ}{dx_{1}}=\frac{Q}{\Omega _{D}~a_{1}a_{2}\cdots a_{D-2}}~V\left(
a_{1},a_{2},a_{3},\cdots ,a_{D-2}\right) \ .
\end{equation}%
The remaining integrations, easily performed sequentially, are given for $%
-a_{1}\leq x_{1}\leq a_{1}$ by 
\begin{gather}
V\left( a_{1},\cdots ,a_{D-2}\right) =2\pi \dint_{-a_{2}\sqrt{%
1-x_{1}^{2}/a_{1}^{2}}}^{+a_{2}\sqrt{1-x_{1}^{2}/a_{1}^{2}}%
}dx_{2}\int_{-a_{3}\sqrt{1-x_{1}^{2}/a_{1}^{2}-x_{2}^{2}/a_{2}^{2}}}^{+a_{3}%
\sqrt{1-x_{1}^{2}/a_{1}^{2}-x_{2}^{2}/a_{2}^{2}}}dx_{3}~\cdots
~\int_{-a_{D-2}\sqrt{1-x_{1}^{2}/a_{1}^{2}-\cdots -x_{D-3}^{2}/a_{D-3}^{2}}%
}^{+a_{D-2}\sqrt{1-x_{1}^{2}/a_{1}^{2}-\cdots -x_{D-3}^{2}/a_{D-3}^{2}}%
}dx_{D-2}  \notag \\
=\Omega _{D-1}~a_{2}a_{3}\cdots a_{D-2}~\left( 1-x_{1}^{2}/a_{1}^{2}\right)
^{\frac{D-3}{2}}\ .
\end{gather}%
Note that $\Omega _{D-1}=\frac{2\pi }{D-3}~\Omega _{D-3}$. \ Thus the final
linear charge density projected along the axis is 
\begin{equation}
\frac{dQ}{dx_{1}}=\frac{Q}{a_{1}}~\frac{\Omega _{D-1}}{\Omega _{D}}\left(
1-x_{1}^{2}/a_{1}^{2}\right) ^{\frac{D-3}{2}}\ ,  \label{LambdaAnyD}
\end{equation}%
for $-a_{1}\leq x_{1}\leq a_{1}$, in agreement with the results in \S 6 of 
\cite{TLC}. \ Clearly, the same form applies for any other principal axis,
i.e. $\frac{dQ}{dx_{k}}=\frac{Q}{a_{k}}\frac{\Omega _{D-1}}{\Omega _{D}}%
\left( 1-x_{k}^{2}/a_{k}^{2}\right) ^{\frac{D-3}{2}}$, for $-a_{k}\leq
x_{k}\leq a_{k}$ and $k=1,\cdots ,D$.

For $D>3$ the non-uniform linear charge density (\ref{LambdaAnyD}) is rather
counter-intuitive as it has a maximum at $x_{1}=0$ and falls monotonically
to zero on either side of the maximum, vanishing at the end points $%
x_{1}=\pm a_{1}$. \ Only for $D=2$ does the result conform to what one would
naively expect for mutually repelling charges placed on a line, namely, a
charge distribution peaked at the ends. \ This is discussed in \cite{TLC}. \ 

As a generalization of (\ref{LambdaAnyD}), consider projecting the charge
onto any subset of the principal axes, for example, onto $x_{1},x_{2},\cdots
,x_{k}$ for $k<D$, as may be accomplished by integrating $\rho _{D}$ over $%
x_{m}$ for $m=k+1,\cdots ,D$. \ Following the same steps as given above, the
result is readily seen to be%
\begin{equation}
\frac{dQ}{dx_{1}dx_{2}\cdots dx_{k}}=\frac{Q}{a_{1}a_{2}\cdots a_{k}}~\frac{%
\Omega _{D-k}}{\Omega _{D}}\left(
1-x_{1}^{2}/a_{1}^{2}-x_{2}^{2}/a_{2}^{2}-\cdots -x_{k}^{2}/a_{k}^{2}\right)
^{\frac{D-k-2}{2}}\ ,  \label{SquashedDensity}
\end{equation}%
for all $x_{1},x_{2},\cdots ,x_{k}$ such that $%
x_{1}^{2}/a_{1}^{2}+x_{2}^{2}/a_{2}^{2}+\cdots +x_{k}^{2}/a_{k}^{2}\leq 1$.
\ Since the final answer here is independent of $a_{k+1},\cdots ,a_{D}$,
this would in fact be the correct charge density on an equipotential $k$%
-dimensional manifold obtained from the original equipotential
hyperellipsoid by letting $a_{m}\rightarrow 0$ for $m=k+1,\cdots ,D$, i.e.
by \textquotedblleft squashing\textquotedblright\ these $D-k$ dimensions. \
In particular, if the original $x_{1}^{2}/a_{1}^{2}+x_{2}^{2}/a_{2}^{2}+%
\cdots +x_{D}^{2}/a_{D}^{2}=1$ hyperellipsoid were completely
\textquotedblleft flattened\textquotedblright\ to a two-dimensional ellipse
embedded in $D$ spatial dimensions, the surface charge density that results
would be%
\begin{equation}
\frac{dQ}{dx_{1}dx_{2}}=\frac{Q}{a_{1}a_{2}}~\frac{\Omega _{D-2}}{\Omega _{D}%
}\left( 1-x_{1}^{2}/a_{1}^{2}-x_{2}^{2}/a_{2}^{2}\right) ^{\frac{D-4}{2}}\ ,
\label{DiskD}
\end{equation}%
for all $x_{1},x_{2}$ such that $x_{1}^{2}/a_{1}^{2}+x_{2}^{2}/a_{2}^{2}\leq
1$. \ Note that this counts the total charge on \emph{both} sides of the
final flattened ellipse, and it nicely generalizes the well-known charge per
area on an ideal, flat, elliptical, equipotential disk embedded in three
dimensions (for example, see \cite{Smythe}), namely,%
\begin{equation}
\frac{dQ}{dA}=\frac{Q}{2\pi a_{1}a_{2}}\frac{1}{\sqrt{%
1-x_{1}^{2}/a_{1}^{2}-x_{2}^{2}/a_{2}^{2}}}\text{ \ \ for \ }D=3\ .
\label{Disk3}
\end{equation}%
For both (\ref{DiskD}) and (\ref{Disk3}), the boundary of the flattened disk
is the ellipse $x_{1}^{2}/a_{1}^{2}+x_{2}^{2}/a_{2}^{2}=1$.

As was the case for the linear charge density in (\ref{LambdaAnyD}), the
result (\ref{SquashedDensity}) is rather counter-intuitive for squashed
manifolds with $D\geq k+2$. \ For $D<k+2$, the charge density $%
dQ/dx_{1}dx_{2}\cdots dx_{k}$ is peaked at the boundary of the squashed
manifold, for which $x_{1}^{2}/a_{1}^{2}+x_{2}^{2}/a_{2}^{2}+\cdots
+x_{k}^{2}/a_{k}^{2}=1$, where the density actually diverges. \ In our
opinion, this would conform with naive expectations for mutually repelling
charges placed on the manifold. \ But for the case $D=k+2$, the charge
density on the squashed manifold is uniform, exactly like that of an ideal
conducting line segment in three dimensions. \ Or, as another particular
case, a flat, elliptical, equipotential disk in four spatial dimensions
would have a constant surface charge density. \ So (\ref{SquashedDensity})
for $D=k+2$ is again non-intuitive, although perhaps it can be reconcilled
with intuition using arguments similar to those advanced for the
equipotential line segment embedded in three dimensions, as discussed in 
\cite{Jackson} and references therein. \ On the other hand, for $D>k+2$, the
charge distribution (\ref{SquashedDensity})\ is peaked at the center of the
squashed manifold and falls monotonically to zero at the boundary where $%
x_{1}^{2}/a_{1}^{2}+x_{2}^{2}/a_{2}^{2}+\cdots +x_{k}^{2}/a_{k}^{2}=1$,
exactly like ideal line segments for $D>3$. \ In our opinion, this is
counter-intuitive. \ Nevertheless, it is what it is. \ 

\section{Potentials and electrostatic fields}

An exact, single-parameter integral expression for the potential surrounding
a static, equipotential, conducting ellipsoid\footnote{%
In this section it will be convenient to use the term \textquotedblleft
ellipsoid\textquotedblright\ rather than \textquotedblleft
hyperellipsoid\textquotedblright\ for any $D$.} carrying a total charge $Q$,
in $D$ spatial dimensions, is in general an elliptic integral, although it
may reduce to an elementary function if some of the $a_{k}$ are equal. \
This fact is well-known for the $D=3$ case \cite{Smythe}. \ Explicitly, if
the charged surface is defined by 
\begin{equation}
\sum_{k=1}^{D}\frac{x_{k}^{2}}{a_{k}^{2}}=1\ ,  \label{EllipsoidD}
\end{equation}%
then the potential is given by 
\begin{equation}
\Phi \left( \overrightarrow{r}\right) =\frac{kQ}{2}~\int_{\Theta \left( 
\overrightarrow{r}\right) }^{\infty }\left( \dprod_{k=1}^{D}\frac{1}{\sqrt{%
a_{k}^{2}+\theta }}\right) d\theta \ ,  \label{PhiAnyD}
\end{equation}%
where the $\Theta $-equipotentials are a set of \emph{confocal} ellipsoids
consisting of all points $\overrightarrow{r}$\ outside the charged ellipsoid
that satisfy, for a given $\Theta $, 
\begin{equation}
\sum_{k=1}^{D}\frac{x_{k}^{2}}{a_{k}^{2}+\Theta }=1\ ,\ \ \ \text{for \ \ }%
\Theta >0\ .  \label{ThetaEllipsoid}
\end{equation}%
Note the charged ellipsoid itself is defined to be at $\Theta =0$. \ If
given an arbitrary point $\overrightarrow{r}$ outside the charged ellipsoid,
to compute the potential at that point it would first be necessary to find
the appropriate $\Theta $ for the given $\overrightarrow{r}$, i.e. find $%
\Theta \left( \overrightarrow{r}\right) $. \ In general, if all the $a_{k}$
are distinct, this would require solving for the appropriate root of the $D$%
th order polynomial in $\Theta $ implicit in (\ref{ThetaEllipsoid}),
something that can always be done in principle (although in practice,
perhaps only numerically, especially if $D>4$ and all the $a_{k}$ are
distinct). \ 

The static electric field is the gradient of the potential, as usual. \ From
(\ref{ThetaEllipsoid}) and (\ref{PhiAnyD}) it follows that%
\begin{equation}
\overrightarrow{\nabla }\Theta \left( \overrightarrow{r}\right) =\left.
\left( \sum_{n=1}^{D}\frac{2x_{n}~\widehat{e_{n}}}{a_{n}^{2}+\Theta }\right)
\right/ \left( \sum_{m=1}^{D}\frac{x_{m}^{2}}{\left( a_{m}^{2}+\Theta
\right) ^{2}}\right) \ ,  \label{GradTheta}
\end{equation}%
\begin{equation}
\overrightarrow{E}\left( \overrightarrow{r}\right) =-\overrightarrow{\nabla }%
\Phi =-\left( \overrightarrow{\nabla }\Theta \right) \frac{d\Phi }{d\Theta }%
=kQ\left. \left( \dprod_{k=1}^{D}\frac{1}{\sqrt{a_{k}^{2}+\Theta }}\right)
\left( \sum_{n=1}^{D}\frac{x_{n}~\widehat{e_{n}}}{a_{n}^{2}+\Theta }\right)
\right/ \left( \sum_{m=1}^{D}\frac{x_{m}^{2}}{\left( a_{m}^{2}+\Theta
\right) ^{2}}\right) \ .  \label{EAnyD}
\end{equation}%
Once again, if only $\overrightarrow{r}$\ is specified, it is necessary to
find $\Theta \left( \overrightarrow{r}\right) $ from (\ref{ThetaEllipsoid})
to evaluate this expression. \ But note that as $r\rightarrow \infty $, it
follows from (\ref{ThetaEllipsoid}) that $\Theta \rightarrow \infty $ with $%
\Theta \underset{r\rightarrow \infty }{\sim }r^{2}$. \ So asymptotically, 
\begin{equation}
\Phi \left( \overrightarrow{r}\right) \underset{r\rightarrow \infty }{\sim }%
\frac{kQ}{D-2}~\frac{1}{r^{D-2}}\ ,\ \ \ \overrightarrow{E}\underset{%
r\rightarrow \infty }{\sim }\frac{kQ~\overrightarrow{r}}{r^{D}}\ .
\end{equation}%
These asymptotic expressions are just the potential and electric field for a
point charge in $D$ spatial dimensions (e.g. see \cite{Sommerfeld}) as
should have been expected.

The direction of the electric field (when multiplied by the sign of $Q$) at
a point on a given $\Theta $-equipotential ellipsoid, is given by the unit
vector%
\begin{equation}
\widehat{E}\left( \overrightarrow{r}\right) =\left. \left( \sum_{n=1}^{D}%
\frac{x_{n}~\widehat{e_{n}}}{a_{n}^{2}+\Theta }\right) \right/ \sqrt{%
\sum_{m=1}^{D}\frac{x_{m}^{2}}{\left( a_{m}^{2}+\Theta \right) ^{2}}}\ .
\label{DirectionEAnyD}
\end{equation}%
As a check, when all the $a_{k}$ are equal, $\widehat{E}=\widehat{r}$, as
expected. \ The (signed) strength of the electric field, at a point on that
same equipotential, is given by%
\begin{equation}
E\left( \overrightarrow{r}\right) =kQ\left. \left( \dprod_{k=1}^{D}\frac{1}{%
\sqrt{a_{k}^{2}+\Theta }}\right) \right/ \sqrt{\sum_{m=1}^{D}\frac{x_{m}^{2}%
}{\left( a_{m}^{2}+\Theta \right) ^{2}}}\ ,  \label{StrengthEAnyD}
\end{equation}%
so that $\overrightarrow{E}\left( \overrightarrow{r}\right) =E\left( 
\overrightarrow{r}\right) ~\widehat{E}\left( \overrightarrow{r}\right) $ for
either sign of $Q$. \ This reduces to $E=kQ/r^{D-1}$\ when all the $a_{k}$
are equal, as expected. \ 

Since the charge is located on the ellipsoid with $\Theta =0$, by
construction, the potential on the charge carrying hypersurface itself is 
\begin{equation}
\Phi \left( \overrightarrow{r}\left( \Theta =0\right) \right) =\frac{kQ}{2}%
~\int_{0}^{\infty }\left( \dprod_{k=1}^{D}\frac{1}{\sqrt{a_{k}^{2}+\theta }}%
\right) d\theta \ .  \label{PhiOnSurface}
\end{equation}%
That is to say, the capacitance of the isolated ellipsoid, defined by $%
Q=C\Phi \left( \overrightarrow{r}\left( \Theta =0\right) \right) $, is also
given by an elliptic integral%
\begin{equation}
C=\frac{2}{k\int_{0}^{\infty }\left( \dprod_{k=1}^{D}\frac{1}{\sqrt{%
a_{k}^{2}+\theta }}\right) d\theta }\ .  \label{Capacitance}
\end{equation}%
Furthermore, the charge density on the ellipsoidal hypersurface may be
obtained directly from Gauss' law, with the normalization for a point charge
determined by%
\begin{equation}
\overrightarrow{\nabla }\cdot \frac{\overrightarrow{r}}{r^{D}}=\Omega
_{D}~\delta ^{D}\left( \overrightarrow{r}\right) \ ,  \label{GaussLaw}
\end{equation}%
where again $\Omega _{D}$ is the total solid angle in $D$ spatial
dimensions. \ Thus the hypersurface charge density is given as usual by the
value of the normal electric field as the hypersurface is approached from
the outside, that is to say by the limit: $k~\Omega _{D}~\sigma _{D}\left( 
\overrightarrow{r}\right) =\lim_{\overrightarrow{r}\rightarrow \text{%
hypersurface}}\widehat{n}\cdot \overrightarrow{E}$ where $\widehat{n}$ is
the outward normal unit vector on the hypersurface. \ For the problem at
hand this limit is just $\lim_{\Theta \rightarrow 0}\widehat{E}\cdot 
\overrightarrow{E}=\left. E\right\vert _{\Theta =0}$. \ So (\ref%
{StrengthEAnyD}) gives the explicit result%
\begin{equation}
\sigma _{D}\left( \overrightarrow{r}\right) =\frac{1}{k~\Omega _{D}}\left.
E\right\vert _{\Theta =0}=\frac{Q}{\Omega _{D}\dprod_{k=1}^{D}a_{k}}\left. 
\frac{1}{\sqrt{\sum_{m=1}^{D}\frac{x_{m}^{2}}{a_{m}^{4}}}}\right\vert
_{\Theta =0}\ ,  \label{SigmaAnyDAgain}
\end{equation}%
thereby confirming both (\ref{RhoAnyD}) and (\ref{SigmaAnyD}).\footnote[4]{%
Using notation that is more consistent with the previous section, $d\Theta
=\left( \overrightarrow{\nabla }\Theta \right) \cdot d\overrightarrow{r}%
=\left\vert \overrightarrow{\nabla }\Theta \right\vert ~\widehat{n}\left(
\Theta \right) \cdot d\overrightarrow{r}=\left\vert \overrightarrow{\nabla }%
\Theta \right\vert ~du\left( \Theta \right) $, where $\widehat{n}\left(
\Theta \right) =\widehat{E}$ is the local normal on the $\Theta $%
-equipotential, and $\left\vert \overrightarrow{\nabla }\Theta \right\vert
=2\left/ \sqrt{\sum_{m=1}^{D}x_{m}^{2}/\left( a_{m}^{2}+\Theta \right) ^{2}}%
\right. $. \ Therefore $\overrightarrow{E}=-\widehat{n}\left( \Theta \right)
~d\Phi /du\left( \Theta \right) =-\widehat{E}~\left\vert \overrightarrow{%
\nabla }\Theta \right\vert ~d\Phi /d\Theta $. \ This is in agreement with (%
\ref{EAnyD}), (\ref{DirectionEAnyD}), and (\ref{StrengthEAnyD}).}

\section{Conclusions}

We have shown that conducting ellipsoids in any number of spatial dimensions
have volume charge densities given by (\ref{RhoAnyD}). \ We are not aware of
any previous literature that gives this particular form for $\rho _{D}$,
even for the case $D=3$.

It is an interesting problem to determine if there are any surface
geometries other than ellipsoids, for conductors of bounded extent, where
the volume charge density has the form (\ref{GenericForm}) with \emph{%
constant} $f$. \ An obvious conjecture is that only ellipsoids have this
property. \ 

As a test of this conjecture, it is a straightforward task to consider \href{https://en.wikipedia.org/wiki/Superegg}%
{supereggs},\ \href{https://en.wikipedia.org/wiki/Superellipsoid}{%
superellipsoids},\ or higher dimensional \textquotedblleft Lam\'{e}
hypersurfaces\textquotedblright\ defined by $\sum_{k=1}^{D}\left(
x_{k}^{2}/a_{k}^{2}\right) ^{N}=1$. \ A preliminary analysis \cite{TLCunpub}
shows that a charge density of the form (\ref{GenericForm}) with constant $f$
fails to give a constant potential within such surfaces unless $N=1$. \
General surfaces remain to be investigated. \ A variety of specific surfaces
with closed-form charge densities may be constructed by methods such as
those in \cite{Singular}.\bigskip 

\noindent \textbf{Acknowledgements:} \ For helpful comments, we thank K.
McDonald (in particular for pointing out \cite{Murphy}) and A. Zangwill (in
particular for pointing out \cite{Singular}). \ This work was supported in
part by a University of Miami Cooper Fellowship.

\section{Appendix: \ Geometry of hyperellipsoids}

For a hyperellipsoid\ embedded in $D$ dimensions, as given by $%
\sum_{k=1}^{D}x_{k}^{2}/a_{k}^{2}=1$, resolving the constraint by solving
for $x_{D}\left( x_{1},\cdots ,x_{D-1}\right) $ gives simple expressions for
the metric, inverse metric, and 2nd fundamental form of the manifold. \ The
results are%
\begin{gather}
g_{km}=\delta _{km}+\frac{a_{D}^{4}}{x_{D}^{2}}\frac{x_{k}x_{m}}{%
a_{k}^{2}a_{m}^{2}}\ ,\ \ \ g^{mn}=\delta _{mn}-\frac{1}{S}\frac{x_{m}x_{n}}{%
a_{m}^{2}a_{n}^{2}}\text{ \ \ for \ \ }k,m,n=1,2,\cdots ,D-1, \\
\   \notag \\
\text{where \ \ }x_{D}^{2}=a_{D}^{2}\left( 1-\sum_{j=1}^{D-1}\frac{x_{j}^{2}%
}{a_{j}^{2}}\right) \ ,\text{ \ \ }\det g_{km}=\frac{a_{D}^{4}}{x_{D}^{2}}%
~S\ ,\ \ \ \text{and \ \ }S\equiv \sum_{j=1}^{D}\frac{x_{j}^{2}}{a_{j}^{4}}%
>0\ ,
\end{gather}%
\begin{eqnarray}
K_{mn} &=&-\widehat{n}\cdot \left( \partial _{m}\partial _{n}\overrightarrow{%
r}\right) =\frac{1}{\sqrt{S}}\frac{1}{a_{m}^{2}}\left( \delta _{mn}+\frac{%
a_{D}^{2}}{a_{n}^{2}}\frac{x_{m}x_{n}}{x_{D}^{2}}\right) \text{ \ \ with\ }%
\overrightarrow{r}\text{ on the manifold \& }\widehat{n}\cdot \partial _{m}%
\overrightarrow{r}=0\ ,  \label{KMatrix} \\
&&\   \notag \\
\mathbb{K}_{kn} &\equiv &g^{km}K_{mn}=\frac{1}{\left( S\right) ^{3/2}}\left( 
\frac{S~\delta _{kn}}{a_{k}a_{n}}+\frac{x_{k}x_{n}}{a_{k}^{2}a_{n}^{2}}%
\left( \frac{1}{a_{D}^{2}}-\frac{1}{a_{n}^{2}}\right) \right) \ ,\text{ \ \
for \ \ }k,m,n=1,\cdots ,D-1.\   \label{BigKMatrix}
\end{eqnarray}%
Note the $\left( \frac{1}{a_{D}^{2}}-\frac{1}{a_{n}^{2}}\right) $ factor in
the matrix $\mathbb{K}_{kn}$ breaks the $k\leftrightarrow n$ symmetry. \
Also note the \emph{coordinate singularity} (as opposed to a physical
singularity) in the metric $g_{mn}$\ and $K_{mn}$\ on the $x_{D}=0$
\textquotedblleft equatorial\textquotedblright\ submanifold. \ However,
there is no such singularity in $\mathbb{K}_{kn}$, whose eigenvalues $\kappa
_{m}$ for $m=1,\cdots ,D-1$, are all \emph{finite} so long as $a_{k}>0$ for $%
k=1,\cdots ,D$. \ The \emph{intrinsic} curvature scalar densities on the
manifold are encoded in%
\begin{equation}
\det \left( 1+\lambda \mathbb{K}\right) =1+\lambda \left(
\sum_{m=1}^{D-1}\kappa _{m}\right) +\lambda ^{2}\left(
\sum_{m>n=1}^{D-1}\kappa _{m}\kappa _{n}\right) +\cdots +\lambda
^{D-1}\left( \dprod_{m=1}^{D-1}\kappa _{m}\right) \ ,
\end{equation}%
where\ $R=\sum_{m>n=1}^{D-1}\kappa _{m}\kappa _{n}$ etc. \ The last term in
the expansion of $\det \left( 1+\lambda \mathbb{K}\right) $ is $\lambda
^{D-1}\det \mathbb{K}$, of course. \ From the above expression (\ref%
{BigKMatrix}) it follows that%
\begin{equation}
\det \mathbb{K}=\frac{1}{\left( \sqrt{S}\right) ^{D+1}}\left(
\dprod_{j=1}^{D}\frac{1}{a_{j}^{2}}\right) \ .
\end{equation}%
Therefore, on a charged, conducting hyperellipsoid embedded in $D$
dimensions, for which the hypersurface charge density is $\sigma _{D}=\frac{1%
}{\Omega _{D}\sqrt{S}}\tprod_{j=1}^{D}\frac{1}{a_{j}}$, it follows that $%
\sigma ^{D+1}\propto \det \mathbb{K}$. \ This\ generalizes the long-known
result for ellipsoids embedded in three dimensions (e.g. see the footnote, p
191, \cite{Kellogg}). \ More precisely 
\begin{equation}
\left( \Omega _{D}\sigma _{D}\right) ^{D+1}=\left( \dprod_{j=1}^{D}\frac{1}{%
a_{j}^{D-1}}\right) \det \mathbb{K=}\left( \dprod_{j=1}^{D}\frac{1}{%
a_{j}^{D-1}}\right) \left( \dprod_{m=1}^{D-1}\kappa _{m}\right) \ .
\end{equation}%
Here $\Omega _{D}=2\pi ^{D/2}/\Gamma \left( D/2\right) $ is the surface area
of a unit radius sphere embedded in $D$ dimensions (i.e. the total
\textquotedblleft solid\textquotedblright\ angle around the center of the
sphere).

\end{document}